\documentclass[osajnl,twocolumn,showpacs,superscriptaddress,10pt]{revtex4-1} 
\usepackage{amsmath,amssymb,graphicx}
\usepackage[colorlinks=true]{hyperref}

\begin{document}

\title{Tuning random lasing in photonic glasses}

\author{Michele Gaio}
\email{Corresponding author: michele.gaio@kcl.ac.uk}
\author{Matilda Peruzzo}
\author{Riccardo Sapienza}

\affiliation{Department of Physics, King's College London, Strand, London WCR
2LS, United Kingdom.}

\begin{abstract}
We present a detailed numerical investigation of the tunability of
a diffusive random laser when Mie resonances are excited. We solve a multimode diffusion model and calculate multiple light scattering in presence of optical gain which includes dispersion in both scattering and gain,
 without any assumptions about the $\beta$ parameter. This allows us to investigate
a realistic photonic glass made of latex spheres and rhodamine and to  quantify
 both the lasing wavelength tunability range and the lasing threshold.
Beyond what is expected by diffusive monochromatic models, the highest threshold is found when the competition between the lasing modes
is strongest and not when the lasing wavelength is furthest from the
maximum of the gain curve.
\end{abstract}

\maketitle

Random lasers (RL) are mirror-less lasing systems which have attracted
a lot of interest due to their structural simplicity. Nowadays they have
been studied in a vast variety of scattering systems ranging from
semiconductor powder  to biological tissue and biocompatible materials \cite{Wiersma2008}. Random lasing originates
from a complex out-of-equilibrium phenomenon with rich multimodes
features \cite{Cao2005} and surprising statistical features \cite{Ghofraniha2015}.
Despite its  potential for practical applications \cite{Wiersma2008},
random lasing technology is still in its infancy with pioneering applications
such as  low coherence light source \cite{Redding2012} and  biosensing \cite{Soraya2015}. One of the factors that has limited practical
applications is the difficulty of controlling the frequency and directionality
of the emission. In conventional lasers the lasing emission can be
tuned by engineering the high finesse cavity which provides the feedback
and thus defines the lasing mode. Instead, feedback in RL 
is provided by multiple scattering and the lasing emission properties
are determined by the complex interplay between gain and losses. Recent
experiments have shown lasing emission controlled  by exploiting
scattering dispersion via resonant scattering sustained by spherical
particles \cite{Gottardo2008,Uppu2011} or by gain dispersion achieved by artificially
increasing absorption in a spectral band \cite{El-dardiry2011}.
Active tuning of the lasing properties has also been achieved by shaping the pump profile to
selectively excite one or a few lasing modes \cite{Bachelard2012,Hisch2013,Bachelard2014}. 

Different theoretical approaches to model random lasing action have
been developed which combine multiple scattering and gain. For uncorrelated
random systems in which interference between the scattered waves can
be neglected, diffusive models are very accurate even in presence
of optical gain \cite{John1996,Wiersma1996} and they provide the
time evolution of the lasing process and a smooth lasing spectrum
with no spiking lasing behaviour \cite{El-Dardiry2012}. The radiative
transport model with gain can also be solved for instance 
with Monte Carlo simulations which consider a random walk of photons
\cite{Balachandran1997,Berger1997} and in which amplification of
single paths can be important in defining the spectral properties
\cite{Mujumdar2004}, and by solving the complete radiative transfer equations \cite{Pierrat2007}.  These approaches allow the study of large systems
($>100$s mean free paths) and geometries similar to real experiments.
Recently, more complete models including field calculations and interference
effects have been developed, based for example on Maxwell-Bloch equations \cite{Tureci2006,Tureci2009}
and the  finite-difference time-domain solution of Maxwell equations in nonlinear media \cite{Jiang2000,Sebbah2002,Conti2007},
but limited to volumes of a few wavelengths cubed. 

In this letter, we investigate
the range of tunability of a diffusive RL with intensity feedback. We solve the diffusion equations  for typical experimental
configurations by including the full
spectral dispersion of both scattering and gain, beyond the stationary
case \cite{John1996} and few modes model \cite{El-Dardiry2012},
and we calculate the full spatial, spectral and lasing dynamics, including mode competition and threshold variation.  

The relevant quantities in a diffusive model are the transport mean
free path $\ell_{t}$, the diffusion constant $D=\ell_{t}v/3$, and
$v$ the speed of energy in the medium, which can all be modulated by resonant scattering \cite{Sapienza2007}. The
optical gain is provided by organic molecules, rhodamine 6G in this
case, described by the stimulated emission cross-section $\sigma_{e}$, the absorption cross section $\sigma_{a}$, 
and the lifetime of the excited state $\tau$. The molecules composing
the gain can be brought to the excited state by an intense pump laser.
Here we label $N_{1}$ the density of molecules in the excited state
and $N$ the total molecule density. We model a translational invariant
slab geometry similar to most experiments with the 1D diffusion equation for the pump $W_{p}(x,t)$
and the emitted light $W^{i}(x,t)$. These equations are coupled to a four level system gain
described by one rate equation for the radiative transition. The set of equation is the following:
\begin{eqnarray}
\frac{\partial N_{1}}{\partial t} & = & \sigma_{a}W_{p}[N-N_{1}]v-\sum_{i}\sigma_{e}^{i}W^{i}N_{1}-\frac{N_{1}}{\tau}\label{eq:odesystem1}\\
\frac{\partial W_{p}}{\partial t} & = & D\frac{\partial^{2}W_{p}}{\partial x^{2}}-\sigma_{a}W_{p}[N-N_{1}]v+\frac{I_{p}}{\ell_{e}}\label{eq:odesystem2}\\
\frac{\partial W^{i}}{\partial t} & = & D\frac{\partial^{2}W_{i}}{\partial x^{2}}+\sigma_{e}^{i}W^{i}N_{1}v+\frac{\phi^{i}}{\tau}N_{1}\label{eq:odesystem3}
\end{eqnarray}
where the emitted light is discretised in spectral bands $\lambda_{i}$, and
$i=1..n$. $I_{p}(x,t)$ is the intensity  temporal profile of
the pump, $\ell_{e}$ is its extinction length and $v$ is the velocity of light in inside the medium. The spontaneous emission (fluorescence) spectrum defines the quantities $\phi^{i}$, with $\sum\phi^{i}=1$. When compared
to previous works \cite{Wiersma1996,El-dardiry2011}, the spontaneous
emission factor $\beta$ is not a free parameter any more, but comes as a solution
to the problem. We solve the system of coupled differential equations
(\ref{eq:odesystem1}-\ref{eq:odesystem3}) by means of standard ODE
solvers implemented in MATLAB. The main advantage of this model is
that it includes spectral dispersion of scattering,
absorption and gain.

\begin{figure}
\includegraphics[scale=0.65]{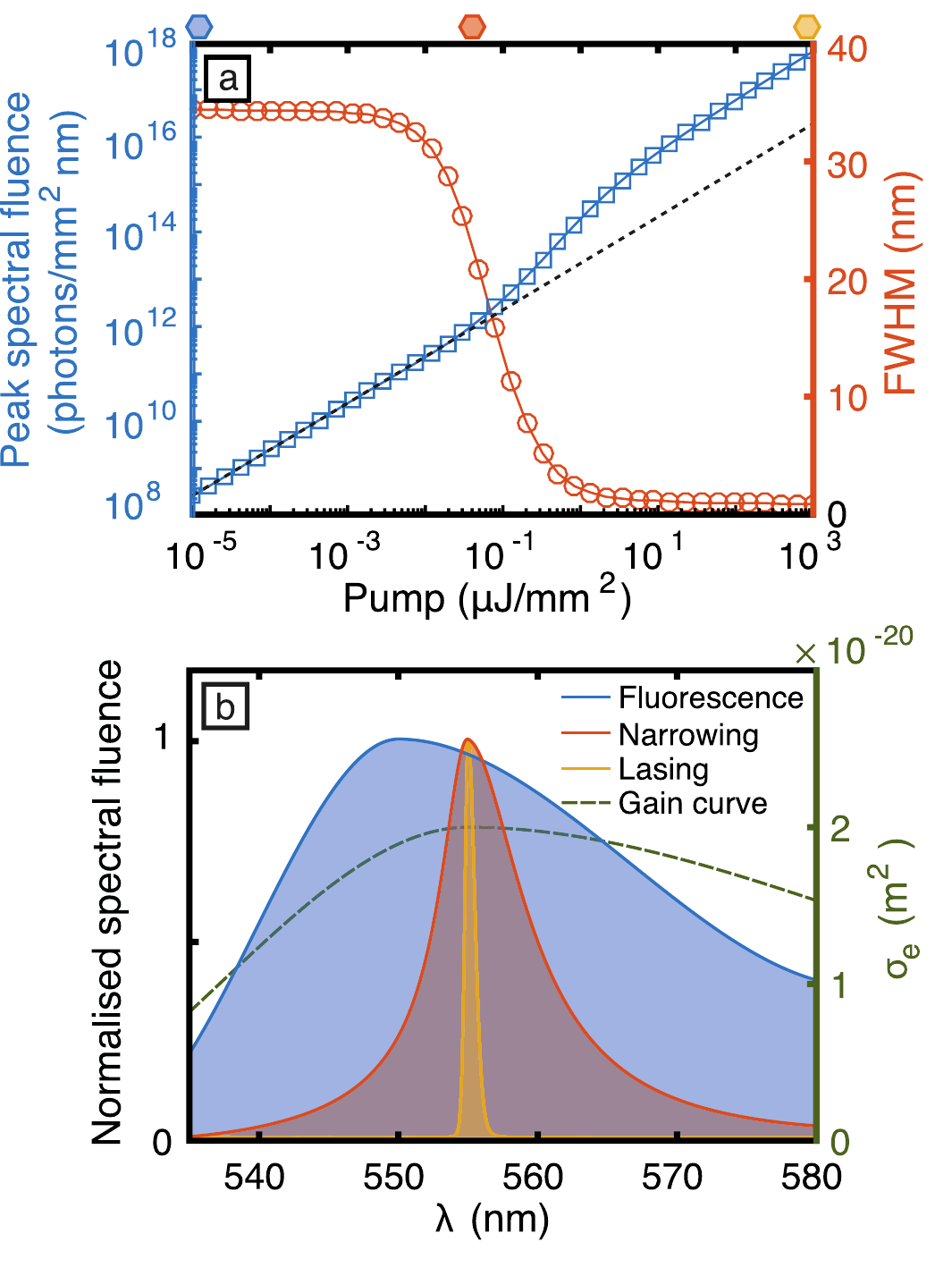}

\caption{(a) Characteristic plot of a random laser: the total peak intensity (blue
squares) and FWHM of the emission spectrum (red circles) are plotted for increasing
pump energy density for a slab system of thickness $L=50~\mu$m, transport
scattering length $\ell_{t}=1.5~\mu$m, and a 1
$m$M concentration of rhodamine 6G providing the gain, excited by a single
pulse of duration $d=6~n$s at 532 nm. The black dashed line is a guide to the eye to highlight
the change of regime from fluorescence to lasing. (b) The emission
spectra (normalised) are reported at different pump intensities: the fluorescence
spectrum (blue) narrows down for increasing pump intensity (red line) down
to a 0.8 $n$m FWHM peak (yellow line). The corresponding pump intensities are colour
coded by the hexagons in panel (a). The lasing occurs at the maximum
of the gain curve (dashed green line).}

\label{fig:figure1}
\end{figure}

We choose to simulate typical experimental conditions \cite{Gottardo2008}:
the sample is a slab 50~$\mu$m thick, with $\ell_{t}$~=~1.5~$\mu$m,
and doped with a concentration of 1mM of rhodamine 6G dye whose emission
properties are taken from ref. \cite{Holzer2000}. The system is pumped
with a 6~ns laser pulse at a wavelength of 532~nm.

Firstly, we consider a conventional RL, such as a polydisperse TiO$_{2}$
powder, where $\ell_{t}$ is not dispersive but roughly constant over
the gain spectrum. In Fig.~\ref{fig:figure1}(a) we plot the peak
spectral fluence diffusing from the system at the air-sample
interface together with the full width half maximum (FWHM) of the
emission spectra as a function of the pump energy density. The onset
of the lasing emission is at $P=$~0.07~mJ/mm$^{2}$, at higher
pump intensities the peak fluence increases super-linearly until gain
saturation is reached; at the same time the emission width quickly
decreases from the broad fluorescence emission to a narrow almost
constant value at saturation. The threshold of the lasing emission
is usually identified in analogy to conventional lasing by considering
the change in slope of the pump-peak emission intensity relation or
by considering the narrowing of the lasing emission. Throughout this
letter we identify the threshold as the half narrowing of the emission
spectrum, which is a typical experimental parameter more accessible
than the change of slope \cite{VanSoest1999}. The FWHM of the emission
narrows from the initial $\sim35$~nm of the rhodamine fluorescence
spectrum to  $\sim0.8$~nm of maximum lasing narrowing. The  $\beta$ parameter can be calculated from Fig.~\ref{fig:figure1}(a) and is  $\beta\simeq0.04$. In Fig.~\ref{fig:figure1}(b)
the emission spectra below and above threshold and for an intermediate
value are shown. Lasing occurs at the maximum of the gain, which is
$\lambda=555.5$~nm. The width of the lasing peak is limited by the
losses through the sample boundaries and by the gain saturation, and,
as we confirmed, it is not affected by the numerical spectral discretisation
which is set to 0.1~nm. This value differs from the typical experimental
values where the final linewidth is typically in the range 5 to 10~nm
\cite{VanSoest1999,Gottardo2008}, as we are not considering any
additional homogeneous and inhomogeneous line-broadening effects.

\begin{figure}
\includegraphics[scale=0.65]{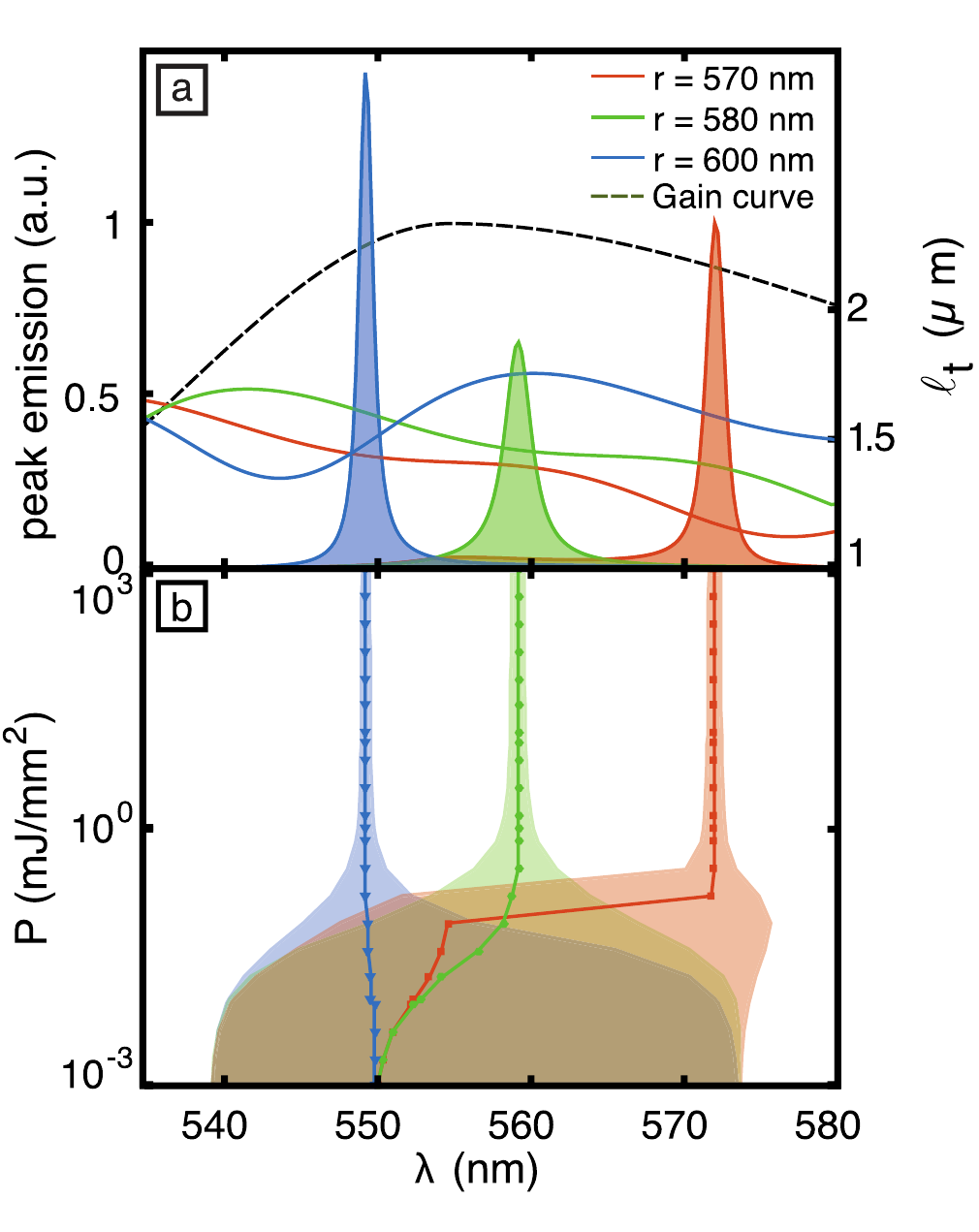}

\caption{The effects of resonant scattering on the lasing spectrum. (a) The Mie resonances
of spherical scatterers ($n=1.6$ in air) of similar size (polydispersty
1\%) are exploited to tune the lasing emission. The emission wavelength
is close to the minimum of the transport mean free path while inside
the gain region (black line). (b) At increasing pump power
the emission peak wavelength shifts from the fluorescence maximum to
the lasing wavelength stabilizing above threshold.}

\label{fig:figure2}
\end{figure}

We consider now the case of a medium characterised by resonant Mie
scattering. We simulate a system composed of close-packed ($n=1.6$, filling
fraction $f=$~0.5) dielectric spheres with 1\% radial polydispersity.
We calculate $\ell_{t}$ by using Mie theory in the
approximation of independent scatterers \cite{craig1983absorption}.
The Mie resonances modulate $\ell_{t}$ as shown by the full lines in Fig.~\ref{fig:figure2}(a), where scattering is plotted together with the
predicted lasing peaks for three sphere diameters. Qualitatively, the lasing frequency
is pulled towards the strongest scattering frequency within the gain
curve. Compared to what is shown in Fig.~\ref{fig:figure1}(b), now
the lasing frequency can be tuned by choosing the Mie resonance.

Fig.~\ref{fig:figure2}(b) shows the narrowing and shifting of the
emission peak while increasing the pump energy. The crossing of the lasing
threshold is now evident as a narrowing of the emission spectrum 
while the lasing peak shifts away from the maximum of the fluorescence
curve. Frequencies close to the maximum of the gain initially dominate,
as they receive a larger fraction of the spontaneous emission, but
those with larger scattering and gain eventually prevail above threshold.

\begin{figure}
\includegraphics[scale=0.65]{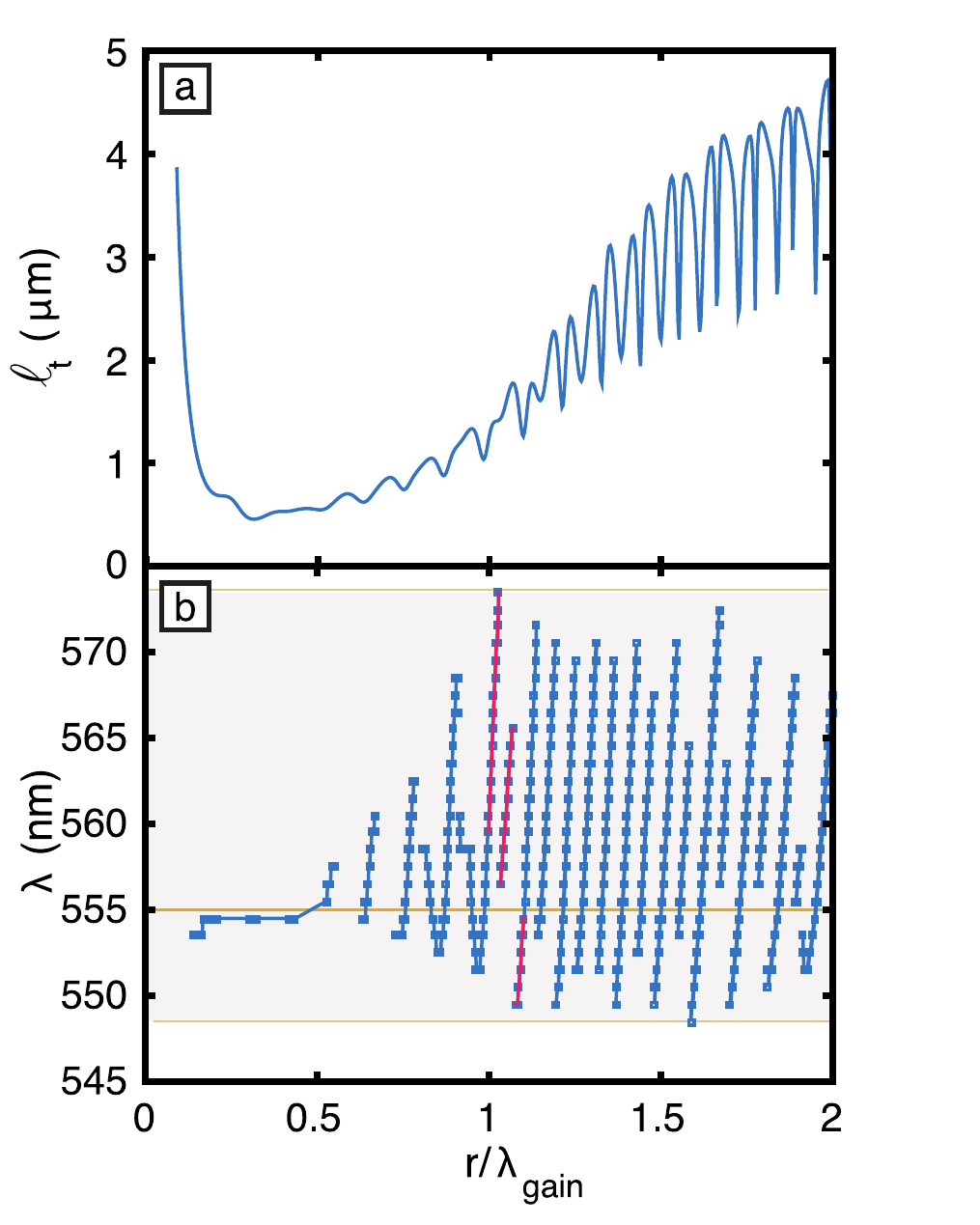}

\caption{Tuning the lasing emission wavelength via resonant scattering.
(a) Transport scattering mean free path $\ell_t$ computed by  Mie scattering (spheres of refractive index $n=1.6$ in air, filling
ratio $f=0.5$ and 1\% polydispersity) at $\lambda_{max}=555.5~n$m.
(b) Lasing emission wavelength for resonant scattering. When the particle size is increased from non-resonant Rayleigh scattering
($r<<\lambda$) to Mie scattering, the lasing emission wavelength is tuned to follow the scattering resonance in a range of 25 $n$m 
inside the gain curve (gray area). The red lines are further plotted in Fig.~\ref{fig:figure4}(a).}

\label{fig:figure3}
\end{figure}

Finally, we study the range of tunability that can be achieved by
resonant scattering. In Fig.~\ref{fig:figure3}(a) we present the
calculated transport mean free path at $\lambda_{gain}=555.5$~nm,
which is the wavelength corresponding to the maximum of the gain.
While for very small scatterers, which are in the Rayleigh regime,
the scattering increases with the particle size, approaching the Mie
regime this trend is reversed. The minimum scattering length is achieved at $r\lesssim\lambda/2$, 
which corresponds to the onset of the first Mie resonances, and subsequently 
$\ell_t$ increases with a resonant behaviour, while the resonances get closer and closer. As previously
shown in Fig.~\ref{fig:figure2}, this modulation in the scattering
induces a tuning of the emission wavelength which is now explicit
in Fig.~\ref{fig:figure3}(b) where we plot the lasing frequency
versus the particle reduced radius $r/\lambda_{gain}$.
While in the Rayleigh regime the lasing emission follows the maximum of the
gain, for $r>\lambda/2$ the emission starts to red-shift following
the Mie resonances.
The wavelength jumps correspond to the appearance of a more favorable
resonance which pulls the lasing wavelength in the blue part of the
gain region. The tunability range for the simulated system is highlighted
by the area in grey and it is roughly $25$~nm.

\begin{figure}
\includegraphics[scale=0.65]{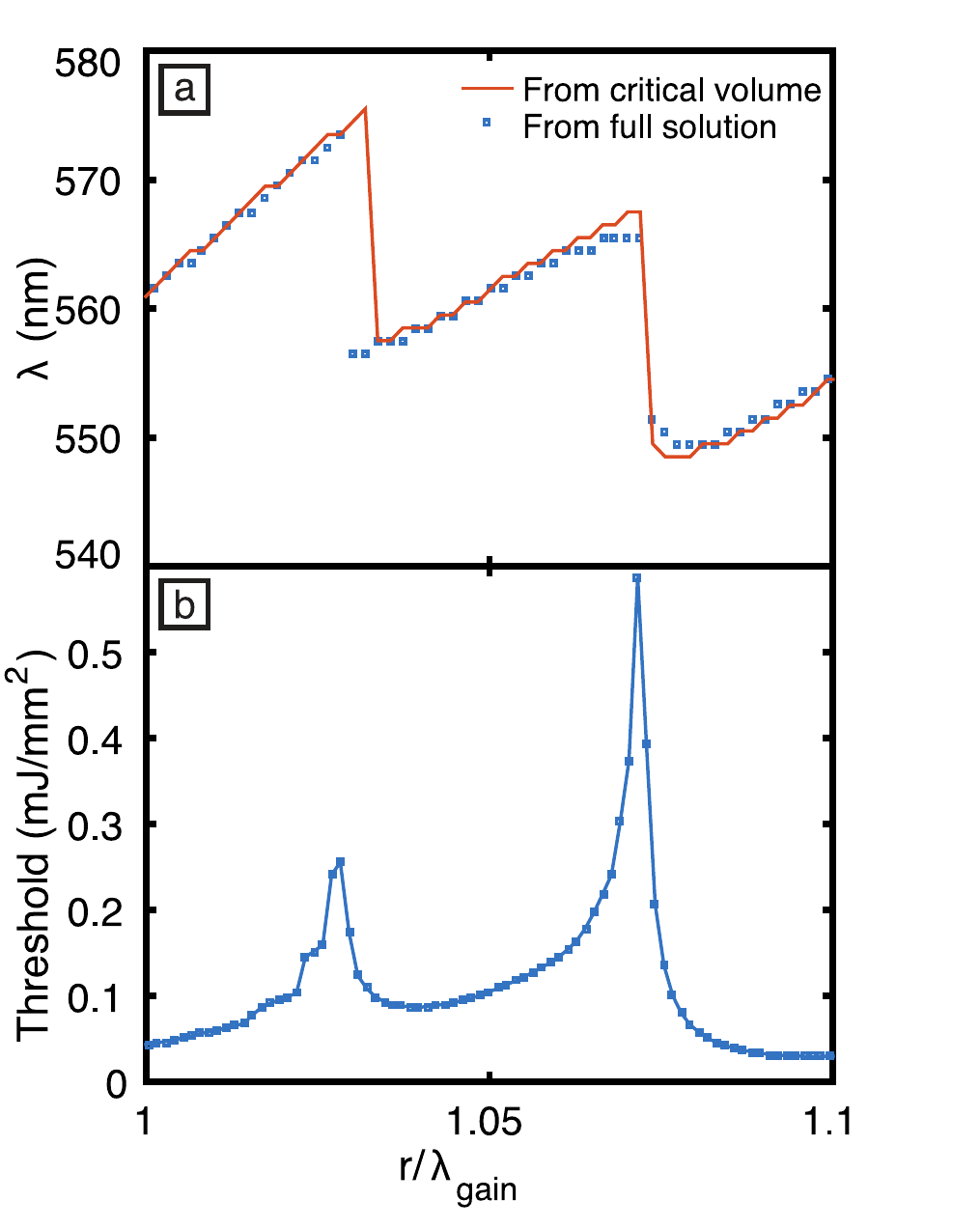}
\caption{Details of the emission wavelength and lasing threshold. (a) Pulled by
the Mie resonances, the lasing emission redshifts  until the next Mie
resonance enters the gain spectral region. This competition is shown by the tooth-saw profile. The numerical values (squares) are
compared to the analytical prediction (orange line) obtained by the minimum
of the critical length $L_{cr}=\pi\sqrt{\ell_{t}\ell_{g}/3}$. (b) The threshold,
defined as the point of half emission narrowing, is quite constant in the tuning range but close to the frequency jumps,
when the competition between the two lasing peaks strongly increases it.}
\label{fig:figure4}
\end{figure}

Fig.~\ref{fig:figure4}(a) is a zoom in of Fig.~\ref{fig:figure3}(b)
and shows in detail the lasing frequency evolution around $r=\lambda$.
The curve consists of three different continuos branches separated
by two jumps. The numerical results of the model are compared to analytical
diffusion calculations which predict lasing occurring at the wavelength which
minimises the critical length $L_{cr}$ \cite{Gottardo2008,Uppu2011}, 
the length above which the gain exceeds the losses. In a slab
geometry, and under the approximation of uniform gain, the critical length
is $L_{cr}=\pi\sqrt{\ell_{t}l_{g}/3}$, where $l_{g}=(N\sigma_{e})^{-1}$.
The minimum of $L_{cr}$ and the numerical results are in good
agreement,  with  significant differences only close to the wavelength jumps where
the simplified analytical model fails to account for the
competition for the gain. We attribute this to the increased
mode competition between two different Mie resonances. Mode competition
is strong for particle sizes around $r=571$~nm and 595.5~nm, and this is reflected by an increased threshold.
For instance, the threshold does not always
decreases monotonically when the lasing emission approaches $\lambda_{gain}$. For the other lasing
frequencies the threshold is quite constant in the whole region of
tunability with values around $0.1$~mJ/mm$^{2}$. 
In addition, we confirm (not shown here) that for particles
size $r=571.9$~nm and $r=595.4$~nm the spectrum develops two competing
peaks of which one finally prevails for large enough pump energy.

In conclusion, a dispersive diffusive gain model is capable to simulate
RL action when driven by gain or scattering resonances. For resonant
scattering we calculated a 25~nm emission tuning
range, with minimal threshold increase. Our dispersive model does not
require any assumption about the $\beta$ parameter and can predict
mode competition and a non-trivial threshold dependence. As a further
extension, the model can include arbitrary absorption curves and dispersion
also for the energy velocity, and can be extended to the full three-dimensional
case at the expense of increased computing time.

We thank Soraya Caixeiro for fruitful discussions. This research was
funded by the Engineering and Physical Sciences Research Council (EPSRC),
 a Leverhulme Trust Research Grant and a FP7 European project.
 The data is publicly available in Figshare \cite{figsharedata}, the code is available on request.

\begin{thebibliography}{27}%
\makeatletter
\providecommand \@ifxundefined [1]{%
 \@ifx{#1\undefined}
}%
\providecommand \@ifnum [1]{%
 \ifnum #1\expandafter \@firstoftwo
 \else \expandafter \@secondoftwo
 \fi
}%
\providecommand \@ifx [1]{%
 \ifx #1\expandafter \@firstoftwo
 \else \expandafter \@secondoftwo
 \fi
}%
\providecommand \natexlab [1]{#1}%
\providecommand \enquote  [1]{``#1''}%
\providecommand \bibnamefont  [1]{#1}%
\providecommand \bibfnamefont [1]{#1}%
\providecommand \citenamefont [1]{#1}%
\providecommand \href@noop [0]{\@secondoftwo}%
\providecommand \href [0]{\begingroup \@sanitize@url \@href}%
\providecommand \@href[1]{\@@startlink{#1}\@@href}%
\providecommand \@@href[1]{\endgroup#1\@@endlink}%
\providecommand \@sanitize@url [0]{\catcode `\\12\catcode `\$12\catcode
  `\&12\catcode `\#12\catcode `\^12\catcode `\_12\catcode `\%12\relax}%
\providecommand \@@startlink[1]{}%
\providecommand \@@endlink[0]{}%
\providecommand \url  [0]{\begingroup\@sanitize@url \@url }%
\providecommand \@url [1]{\endgroup\@href {#1}{\urlprefix }}%
\providecommand \urlprefix  [0]{URL }%
\providecommand \Eprint [0]{\href }%
\providecommand \doibase [0]{http://dx.doi.org/}%
\providecommand \selectlanguage [0]{\@gobble}%
\providecommand \bibinfo  [0]{\@secondoftwo}%
\providecommand \bibfield  [0]{\@secondoftwo}%
\providecommand \translation [1]{[#1]}%
\providecommand \BibitemOpen [0]{}%
\providecommand \bibitemStop [0]{}%
\providecommand \bibitemNoStop [0]{.\EOS\space}%
\providecommand \EOS [0]{\spacefactor3000\relax}%
\providecommand \BibitemShut  [1]{\csname bibitem#1\endcsname}%
\let\auto@bib@innerbib\@empty
\bibitem [{\citenamefont {Wiersma}(2008)}]{Wiersma2008}%
  \BibitemOpen
  \bibfield  {author} {\bibinfo {author} {\bibfnamefont {D.}~\bibnamefont
  {Wiersma}},\ }\href
  {http://www.nature.com/nphys/journal/v4/n5/abs/nphys971.html} {\bibfield
  {journal} {\bibinfo  {journal} {Nat. Phys.}\ }\textbf {\bibinfo {volume} {4}}
  (\bibinfo {year} {2008})}\BibitemShut {NoStop}%
\bibitem [{\citenamefont {Cao}(2006)}]{Cao2005}%
  \BibitemOpen
  \bibfield  {author} {\bibinfo {author} {\bibfnamefont {H.}~\bibnamefont
  {Cao}},\ }\href {http://stacks.iop.org/0305-4470/39/i=2/a=C01} {\bibfield
  {journal} {\bibinfo  {journal} {Journal of Physics A: Mathematical and
  General}\ }\textbf {\bibinfo {volume} {39}},\ \bibinfo {pages} {467}
  (\bibinfo {year} {2006})}\BibitemShut {NoStop}%
\bibitem [{\citenamefont {Ghofraniha}\ \emph {et~al.}(2015)\citenamefont
  {Ghofraniha}, \citenamefont {Viola}, \citenamefont {{Di Maria}},
  \citenamefont {Barbarella}, \citenamefont {Gigli}, \citenamefont {Leuzzi},\
  and\ \citenamefont {Conti}}]{Ghofraniha2015}%
  \BibitemOpen
  \bibfield  {author} {\bibinfo {author} {\bibfnamefont {N.}~\bibnamefont
  {Ghofraniha}}, \bibinfo {author} {\bibfnamefont {I.}~\bibnamefont {Viola}},
  \bibinfo {author} {\bibfnamefont {F.}~\bibnamefont {{Di Maria}}}, \bibinfo
  {author} {\bibfnamefont {G.}~\bibnamefont {Barbarella}}, \bibinfo {author}
  {\bibfnamefont {G.}~\bibnamefont {Gigli}}, \bibinfo {author} {\bibfnamefont
  {L.}~\bibnamefont {Leuzzi}}, \ and\ \bibinfo {author} {\bibfnamefont
  {C.}~\bibnamefont {Conti}},\ }\href {http://dx.doi.org/10.1038/ncomms7058
  10.1038/ncomms7058} {\bibfield  {journal} {\bibinfo  {journal} {Nat Commun}\
  } (\bibinfo {year} {2015})}\BibitemShut {NoStop}%
\bibitem [{\citenamefont {Redding}\ \emph {et~al.}(2012)\citenamefont
  {Redding}, \citenamefont {Choma},\ and\ \citenamefont {Cao}}]{Redding2012}%
  \BibitemOpen
  \bibfield  {author} {\bibinfo {author} {\bibfnamefont {B.}~\bibnamefont
  {Redding}}, \bibinfo {author} {\bibfnamefont {M.~A.}\ \bibnamefont {Choma}},
  \ and\ \bibinfo {author} {\bibfnamefont {H.}~\bibnamefont {Cao}},\ }\href
  {http://dx.doi.org/10.1038/nphoton.2012.90
  http://www.nature.com/nphoton/journal/v6/n6/abs/nphoton.2012.90.html\#supplementary-information}
  {\bibfield  {journal} {\bibinfo  {journal} {Nat Phot.}\ }\textbf {\bibinfo
  {volume} {6}},\ \bibinfo {pages} {355} (\bibinfo {year} {2012})}\BibitemShut
  {NoStop}%
\bibitem [{\citenamefont {Caixeiro}\ \emph {et~al.}(2015)\citenamefont
  {Caixeiro}, \citenamefont {Gaio}, \citenamefont {Marelli}, \citenamefont
  {Omenetto},\ and\ \citenamefont {Sapienza}}]{Soraya2015}%
  \BibitemOpen
  \bibfield  {author} {\bibinfo {author} {\bibfnamefont {S.}~\bibnamefont
  {Caixeiro}}, \bibinfo {author} {\bibfnamefont {M.}~\bibnamefont {Gaio}},
  \bibinfo {author} {\bibfnamefont {Marelli}~\bibnamefont {B.}}, \bibinfo {author}
  {\bibfnamefont {F.~G.}~\bibnamefont {Omenetto}}, \ and\ \bibinfo {author}
  {\bibfnamefont {R.}~\bibnamefont {Sapienza}},\ }\href@noop {} {\bibfield
  {journal} {\bibinfo  {journal} {In pubblication}\ } (\bibinfo {year}
  {2015})}\BibitemShut {NoStop}%
\bibitem [{\citenamefont {Gottardo}\ \emph {et~al.}(2008)\citenamefont
  {Gottardo}, \citenamefont {Sapienza}, \citenamefont {Garcia}, \citenamefont
  {Blanco}, \citenamefont {Wiersma},\ and\ \citenamefont
  {Lopez}}]{Gottardo2008}%
  \BibitemOpen
  \bibfield  {author} {\bibinfo {author} {\bibfnamefont {S.}~\bibnamefont
  {Gottardo}}, \bibinfo {author} {\bibfnamefont {R.}~\bibnamefont {Sapienza}},
  \bibinfo {author} {\bibfnamefont {P.~D.}\ \bibnamefont {Garcia}}, \bibinfo
  {author} {\bibfnamefont {A.}~\bibnamefont {Blanco}}, \bibinfo {author}
  {\bibfnamefont {D.~S.}\ \bibnamefont {Wiersma}}, \ and\ \bibinfo {author}
  {\bibfnamefont {C.}~\bibnamefont {Lopez}},\ }\href
  {http://dx.doi.org/10.1038/nphoton.2008.102
  http://www.nature.com/nphoton/journal/v2/n7/suppinfo/nphoton.2008.102\_S1.html}
  {\bibfield  {journal} {\bibinfo  {journal} {Nat Phot.}\ }\textbf {\bibinfo
  {volume} {2}},\ \bibinfo {pages} {429} (\bibinfo {year} {2008})}\BibitemShut
  {NoStop}%
\bibitem [{\citenamefont {Uppu}\ and\ \citenamefont
  {Mujumdar}(2011)}]{Uppu2011}%
  \BibitemOpen
  \bibfield  {author} {\bibinfo {author} {\bibfnamefont {R.}~\bibnamefont
  {Uppu}}\ and\ \bibinfo {author} {\bibfnamefont {S.}~\bibnamefont
  {Mujumdar}},\ }\href {\doibase 10.1364/OE.19.023523} {\bibfield  {journal}
  {\bibinfo  {journal} {Opt. Express}\ }\textbf {\bibinfo {volume} {19}},\
  \bibinfo {pages} {23523} (\bibinfo {year} {2011})}\BibitemShut {NoStop}%
\bibitem [{\citenamefont {El-Dardiry}\ and\ \citenamefont
  {Lagendijk}(2011)}]{El-dardiry2011}%
  \BibitemOpen
  \bibfield  {author} {\bibinfo {author} {\bibfnamefont {R.~G.}\ \bibnamefont
  {El-Dardiry}}\ and\ \bibinfo {author} {\bibfnamefont {A.}~\bibnamefont
  {Lagendijk}},\ }\href {\doibase 10.1063/1.3571452} {\bibfield  {journal}
  {\bibinfo  {journal} {Applied Physics Letters}\ }\textbf {\bibinfo {volume}
  {98}},\ \bibinfo {pages} {161106} (\bibinfo {year} {2011})}\BibitemShut
  {NoStop}%
\bibitem [{\citenamefont {Bachelard}\ \emph {et~al.}(2012)\citenamefont
  {Bachelard}, \citenamefont {Andreasen}, \citenamefont {Gigan},\ and\
  \citenamefont {Sebbah}}]{Bachelard2012}%
  \BibitemOpen
  \bibfield  {author} {\bibinfo {author} {\bibfnamefont {N.}~\bibnamefont
  {Bachelard}}, \bibinfo {author} {\bibfnamefont {J.}~\bibnamefont
  {Andreasen}}, \bibinfo {author} {\bibfnamefont {S.}~\bibnamefont {Gigan}}, \
  and\ \bibinfo {author} {\bibfnamefont {P.}~\bibnamefont {Sebbah}},\ }\href
  {\doibase 10.1103/PhysRevLett.109.033903} {\bibfield  {journal} {\bibinfo
  {journal} {Phys. Rev. Lett.}\ }\textbf {\bibinfo {volume} {109}},\ \bibinfo
  {pages} {033903} (\bibinfo {year} {2012})}\BibitemShut {NoStop}%
\bibitem [{\citenamefont {Hisch}\ \emph {et~al.}(2013)\citenamefont {Hisch},
  \citenamefont {Liertzer}, \citenamefont {Pogany}, \citenamefont {Mintert},\
  and\ \citenamefont {Rotter}}]{Hisch2013}%
  \BibitemOpen
  \bibfield  {author} {\bibinfo {author} {\bibfnamefont {T.}~\bibnamefont
  {Hisch}}, \bibinfo {author} {\bibfnamefont {M.}~\bibnamefont {Liertzer}},
  \bibinfo {author} {\bibfnamefont {D.}~\bibnamefont {Pogany}}, \bibinfo
  {author} {\bibfnamefont {F.}~\bibnamefont {Mintert}}, \ and\ \bibinfo
  {author} {\bibfnamefont {S.}~\bibnamefont {Rotter}},\ }\href {\doibase
  10.1103/PhysRevLett.111.023902} {\bibfield  {journal} {\bibinfo  {journal}
  {Phys. Rev. Lett.}\ }\textbf {\bibinfo {volume} {111}},\ \bibinfo {pages}
  {023902} (\bibinfo {year} {2013})}\BibitemShut {NoStop}%
\bibitem [{\citenamefont {Bachelard}\ \emph {et~al.}(2014)\citenamefont
  {Bachelard}, \citenamefont {Gigan}, \citenamefont {Noblin},\ and\
  \citenamefont {Sebbah}}]{Bachelard2014}%
  \BibitemOpen
  \bibfield  {author} {\bibinfo {author} {\bibfnamefont {N.}~\bibnamefont
  {Bachelard}}, \bibinfo {author} {\bibfnamefont {S.}~\bibnamefont {Gigan}},
  \bibinfo {author} {\bibfnamefont {X.}~\bibnamefont {Noblin}}, \ and\ \bibinfo
  {author} {\bibfnamefont {P.}~\bibnamefont {Sebbah}},\ }\href {\doibase
  10.1038/nphys2939} {\bibfield  {journal} {\bibinfo  {journal} {Nat. Phys.}\
  }\textbf {\bibinfo {volume} {10}},\ \bibinfo {pages} {426} (\bibinfo {year}
  {2014})}\BibitemShut {NoStop}%
\bibitem [{\citenamefont {John}\ and\ \citenamefont {Pang}(1996)}]{John1996}%
  \BibitemOpen
  \bibfield  {author} {\bibinfo {author} {\bibfnamefont {S.}~\bibnamefont
  {John}}\ and\ \bibinfo {author} {\bibfnamefont {G.}~\bibnamefont {Pang}},\
  }\href {\doibase 10.1103/PhysRevA.54.3642} {\bibfield  {journal} {\bibinfo
  {journal} {Phys. Rev. A}\ }\textbf {\bibinfo {volume} {54}},\ \bibinfo
  {pages} {3642} (\bibinfo {year} {1996})}\BibitemShut {NoStop}%
\bibitem [{\citenamefont {Wiersma}\ and\ \citenamefont
  {Lagendijk}(1996)}]{Wiersma1996}%
  \BibitemOpen
  \bibfield  {author} {\bibinfo {author} {\bibfnamefont {D.~S.}\ \bibnamefont
  {Wiersma}}\ and\ \bibinfo {author} {\bibfnamefont {A.}~\bibnamefont
  {Lagendijk}},\ }\href {\doibase 10.1103/PhysRevE.54.4256} {\bibfield
  {journal} {\bibinfo  {journal} {Phys. Rev. E}\ }\textbf {\bibinfo {volume}
  {54}},\ \bibinfo {pages} {4256} (\bibinfo {year} {1996})}\BibitemShut
  {NoStop}%
\bibitem [{\citenamefont {El-Dardiry}\ \emph {et~al.}(2012)\citenamefont
  {El-Dardiry}, \citenamefont {Mooiweer},\ and\ \citenamefont
  {Lagendijk}}]{El-Dardiry2012}%
  \BibitemOpen
  \bibfield  {author} {\bibinfo {author} {\bibfnamefont {R.~G.~S.}\
  \bibnamefont {El-Dardiry}}, \bibinfo {author} {\bibfnamefont
  {R.}~\bibnamefont {Mooiweer}}, \ and\ \bibinfo {author} {\bibfnamefont
  {A.}~\bibnamefont {Lagendijk}},\ }\href
  {http://stacks.iop.org/1367-2630/14/i=11/a=113031} {\bibfield  {journal}
  {\bibinfo  {journal} {New Journal of Physics}\ }\textbf {\bibinfo {volume}
  {14}},\ \bibinfo {pages} {113031} (\bibinfo {year} {2012})}\BibitemShut
  {NoStop}%
\bibitem [{\citenamefont {Balachandran}\ \emph {et~al.}(1997)\citenamefont
  {Balachandran}, \citenamefont {Lawandy},\ and\ \citenamefont
  {Moon}}]{Balachandran1997}%
  \BibitemOpen
  \bibfield  {author} {\bibinfo {author} {\bibfnamefont {R.~M.}\ \bibnamefont
  {Balachandran}}, \bibinfo {author} {\bibfnamefont {N.~M.}\ \bibnamefont
  {Lawandy}}, \ and\ \bibinfo {author} {\bibfnamefont {J.~A.}\ \bibnamefont
  {Moon}},\ }\href {\doibase 10.1364/OL.22.000319} {\bibfield  {journal}
  {\bibinfo  {journal} {Opt. Lett.}\ }\textbf {\bibinfo {volume} {22}},\
  \bibinfo {pages} {319} (\bibinfo {year} {1997})}\BibitemShut {NoStop}%
\bibitem [{\citenamefont {Berger}\ \emph {et~al.}(1997)\citenamefont {Berger},
  \citenamefont {Kempe},\ and\ \citenamefont {Genack}}]{Berger1997}%
  \BibitemOpen
  \bibfield  {author} {\bibinfo {author} {\bibfnamefont {G.~A.}\ \bibnamefont
  {Berger}}, \bibinfo {author} {\bibfnamefont {M.}~\bibnamefont {Kempe}}, \
  and\ \bibinfo {author} {\bibfnamefont {A.~Z.}\ \bibnamefont {Genack}},\
  }\href {\doibase 10.1103/PhysRevE.56.6118} {\bibfield  {journal} {\bibinfo
  {journal} {Phys. Rev. E}\ }\textbf {\bibinfo {volume} {56}},\ \bibinfo
  {pages} {6118} (\bibinfo {year} {1997})}\BibitemShut {NoStop}%
\bibitem [{\citenamefont {Mujumdar}\ \emph {et~al.}(2004)\citenamefont
  {Mujumdar}, \citenamefont {Ricci}, \citenamefont {Torre},\ and\ \citenamefont
  {Wiersma}}]{Mujumdar2004}%
  \BibitemOpen
  \bibfield  {author} {\bibinfo {author} {\bibfnamefont {S.}~\bibnamefont
  {Mujumdar}}, \bibinfo {author} {\bibfnamefont {M.}~\bibnamefont {Ricci}},
  \bibinfo {author} {\bibfnamefont {R.}~\bibnamefont {Torre}}, \ and\ \bibinfo
  {author} {\bibfnamefont {D.~S.}\ \bibnamefont {Wiersma}},\ }\href {\doibase
  10.1103/PhysRevLett.93.053903} {\bibfield  {journal} {\bibinfo  {journal}
  {Phys. Rev. Lett.}\ }\textbf {\bibinfo {volume} {93}},\ \bibinfo {pages}
  {053903} (\bibinfo {year} {2004})}\BibitemShut {NoStop}%
\bibitem [{\citenamefont {Pierrat}\ and\ \citenamefont
  {Carminati}(2007)}]{Pierrat2007}%
  \BibitemOpen
  \bibfield  {author} {\bibinfo {author} {\bibfnamefont {R.}~\bibnamefont
  {Pierrat}}\ and\ \bibinfo {author} {\bibfnamefont {R.}~\bibnamefont
  {Carminati}},\ }\href {\doibase 10.1103/PhysRevA.76.023821} {\bibfield
  {journal} {\bibinfo  {journal} {Phys. Rev. A}\ }\textbf {\bibinfo {volume}
  {76}},\ \bibinfo {pages} {023821} (\bibinfo {year} {2007})}\BibitemShut
  {NoStop}%
\bibitem [{\citenamefont {T\"ureci}\ \emph {et~al.}(2006)\citenamefont
  {T\"ureci}, \citenamefont {Stone},\ and\ \citenamefont
  {Collier}}]{Tureci2006}%
  \BibitemOpen
  \bibfield  {author} {\bibinfo {author} {\bibfnamefont {H.~E.}\ \bibnamefont
  {T\"ureci}}, \bibinfo {author} {\bibfnamefont {A.~D.}\ \bibnamefont {Stone}},
  \ and\ \bibinfo {author} {\bibfnamefont {B.}~\bibnamefont {Collier}},\ }\href
  {\doibase 10.1103/PhysRevA.74.043822} {\bibfield  {journal} {\bibinfo
  {journal} {Phys. Rev. A}\ }\textbf {\bibinfo {volume} {74}},\ \bibinfo
  {pages} {043822} (\bibinfo {year} {2006})}\BibitemShut {NoStop}%
\bibitem [{\citenamefont {Tureci}\ \emph {et~al.}(2009)\citenamefont {Tureci},
  \citenamefont {Stone}, \citenamefont {Ge}, \citenamefont {Rotter},\ and\
  \citenamefont {Tandy}}]{Tureci2009}%
  \BibitemOpen
  \bibfield  {author} {\bibinfo {author} {\bibfnamefont {H.~E.}\ \bibnamefont
  {Tureci}}, \bibinfo {author} {\bibfnamefont {A.~D.}\ \bibnamefont {Stone}},
  \bibinfo {author} {\bibfnamefont {L.}~\bibnamefont {Ge}}, \bibinfo {author}
  {\bibfnamefont {S.}~\bibnamefont {Rotter}}, \ and\ \bibinfo {author}
  {\bibfnamefont {R.~J.}\ \bibnamefont {Tandy}},\ }\href
  {http://stacks.iop.org/0951-7715/22/i=1/a=C01} {\bibfield  {journal}
  {\bibinfo  {journal} {Nonlinearity}\ }\textbf {\bibinfo {volume} {22}},\
  \bibinfo {pages} {C1} (\bibinfo {year} {2009})}\BibitemShut {NoStop}%
\bibitem [{\citenamefont {Jiang}\ and\ \citenamefont
  {Soukoulis}(2000)}]{Jiang2000}%
  \BibitemOpen
  \bibfield  {author} {\bibinfo {author} {\bibfnamefont {X.}~\bibnamefont
  {Jiang}}\ and\ \bibinfo {author} {\bibfnamefont {C.~M.}\ \bibnamefont
  {Soukoulis}},\ }\href {\doibase 10.1103/PhysRevLett.85.70} {\bibfield
  {journal} {\bibinfo  {journal} {Phys. Rev. Lett.}\ }\textbf {\bibinfo
  {volume} {85}},\ \bibinfo {pages} {70} (\bibinfo {year} {2000})}\BibitemShut
  {NoStop}%
\bibitem [{\citenamefont {Sebbah}\ and\ \citenamefont
  {Vanneste}(2002)}]{Sebbah2002}%
  \BibitemOpen
  \bibfield  {author} {\bibinfo {author} {\bibfnamefont {P.}~\bibnamefont
  {Sebbah}}\ and\ \bibinfo {author} {\bibfnamefont {C.}~\bibnamefont
  {Vanneste}},\ }\href {\doibase 10.1103/PhysRevB.66.144202} {\bibfield
  {journal} {\bibinfo  {journal} {Phys. Rev. B}\ }\textbf {\bibinfo {volume}
  {66}},\ \bibinfo {pages} {144202} (\bibinfo {year} {2002})}\BibitemShut
  {NoStop}%
\bibitem [{\citenamefont {Conti}\ \emph {et~al.}(2007)\citenamefont {Conti},
  \citenamefont {Angelani},\ and\ \citenamefont {Ruocco}}]{Conti2007}%
  \BibitemOpen
  \bibfield  {author} {\bibinfo {author} {\bibfnamefont {C.}~\bibnamefont
  {Conti}}, \bibinfo {author} {\bibfnamefont {L.}~\bibnamefont {Angelani}}, \
  and\ \bibinfo {author} {\bibfnamefont {G.}~\bibnamefont {Ruocco}},\ }\href
  {\doibase 10.1103/PhysRevA.75.033812} {\bibfield  {journal} {\bibinfo
  {journal} {Phys. Rev. A}\ }\textbf {\bibinfo {volume} {75}},\ \bibinfo
  {pages} {033812} (\bibinfo {year} {2007})}\BibitemShut {NoStop}%
\bibitem [{\citenamefont {Sapienza}\ \emph {et~al.}(2007)\citenamefont
  {Sapienza}, \citenamefont {Garc{\'i}a}, \citenamefont {Bertolotti},
  \citenamefont {Mart{\'i}n}, \citenamefont {Blanco}, \citenamefont {Vi{\~n}a},
  \citenamefont {L{\'o}pez},\ and\ \citenamefont {Wiersma}}]{Sapienza2007}%
  \BibitemOpen
  \bibfield  {author} {\bibinfo {author} {\bibfnamefont {R.}~\bibnamefont
  {Sapienza}}, \bibinfo {author} {\bibfnamefont {P.~D.}\ \bibnamefont
  {Garc{\'i}a}}, \bibinfo {author} {\bibfnamefont {J.}~\bibnamefont
  {Bertolotti}}, \bibinfo {author} {\bibfnamefont {M.~D.}\ \bibnamefont
  {Mart{\'i}n}}, \bibinfo {author} {\bibfnamefont {{\'A}.}~\bibnamefont
  {Blanco}}, \bibinfo {author} {\bibfnamefont {L.}~\bibnamefont {Vi{\~n}a}},
  \bibinfo {author} {\bibfnamefont {C.}~\bibnamefont {L{\'o}pez}}, \ and\
  \bibinfo {author} {\bibfnamefont {D.~S.}\ \bibnamefont {Wiersma}},\ }\href
  {\doibase 10.1103/PhysRevLett.99.233902} {\bibfield  {journal} {\bibinfo
  {journal} {Phys. Rev. Lett.}\ }\textbf {\bibinfo {volume} {99}},\ \bibinfo
  {pages} {233902} (\bibinfo {year} {2007})}\BibitemShut {NoStop}%
\bibitem [{\citenamefont {Holzer}\ \emph {et~al.}(0000)\citenamefont {Holzer},
  \citenamefont {Gratz}, \citenamefont {Schmitt}, \citenamefont {Penzkofer},
  \citenamefont {Costela}, \citenamefont {Garca-Moreno}, \citenamefont
  {Sastre},\ and\ \citenamefont {Duarte}}]{Holzer2000}%
  \BibitemOpen
  \bibfield  {author} {\bibinfo {author} {\bibfnamefont {W.}~\bibnamefont
  {Holzer}}, \bibinfo {author} {\bibfnamefont {H.}~\bibnamefont {Gratz}},
  \bibinfo {author} {\bibfnamefont {T.}~\bibnamefont {Schmitt}}, \bibinfo
  {author} {\bibfnamefont {A.}~\bibnamefont {Penzkofer}}, \bibinfo {author}
  {\bibfnamefont {A.}~\bibnamefont {Costela}}, \bibinfo {author} {\bibfnamefont
  {I.}~\bibnamefont {Garca-Moreno}}, \bibinfo {author} {\bibfnamefont
  {R.}~\bibnamefont {Sastre}}, \ and\ \bibinfo {author} {\bibfnamefont
  {F.}~\bibnamefont {Duarte}},\ }\href {\doibase
  doi:10.1016/S0301-0104(00)00101-4} {\bibfield  {journal} {\bibinfo  {journal}
  {Chemical Physics}\ }\textbf {\bibinfo {volume} {256}},\ \bibinfo {pages}
  {125} (\bibinfo {year} {2000-05-15T00:00:00})}\BibitemShut {NoStop}%
\bibitem [{\citenamefont {van Soest}\ \emph {et~al.}(1999)\citenamefont {van
  Soest}, \citenamefont {Tomita},\ and\ \citenamefont
  {Lagendijk}}]{VanSoest1999}%
  \BibitemOpen
  \bibfield  {author} {\bibinfo {author} {\bibfnamefont {G.}~\bibnamefont {van
  Soest}}, \bibinfo {author} {\bibfnamefont {M.}~\bibnamefont {Tomita}}, \ and\
  \bibinfo {author} {\bibfnamefont {A.}~\bibnamefont {Lagendijk}},\ }\href
  {\doibase 10.1364/OL.24.000306} {\bibfield  {journal} {\bibinfo  {journal}
  {Opt. Lett.}\ }\textbf {\bibinfo {volume} {24}},\ \bibinfo {pages} {306}
  (\bibinfo {year} {1999})}\BibitemShut {NoStop}%
\bibitem [{\citenamefont {Bohren}\ and\ \citenamefont
  {Huffman}(1983)}]{craig1983absorption}%
  \BibitemOpen
  \bibfield  {author} {\bibinfo {author} {\bibfnamefont {C.~F.}\ \bibnamefont
  {Bohren}}\ and\ \bibinfo {author} {\bibfnamefont {D.}~\bibnamefont
  {Huffman}},\ }\href {http://books.google.co.uk/books?id=S1RCZ8BjgN0C} {\emph
  {\bibinfo {title} {Absorption and scattering of light by small particles}}},\
  Wiley science paperback series\ (\bibinfo  {publisher} {Wiley},\ \bibinfo
  {year} {1983})\BibitemShut {NoStop}%
 \bibitem [{\citenamefont {Gaio}\\emph{et~al.}(2015)}]{figsharedata}%
  \BibitemOpen
  \bibfield  {author} {\bibinfo {author} {\bibfnamefont {M.}~\bibnamefont
  {Gaio}},\ \bibinfo {author} {\bibfnamefont {M.}~\bibnamefont
  {Peruzzo}},\ and\ \bibinfo {author} {\bibfnamefont {R.}~\bibnamefont
  {Sapienza}},\ }\href {\doibase doi:10.6084/m9.figshare.1301143}{\,
  \ doi:~\bibinfo  {doi} {10.6084/m9.figshare.1301143}}\ (\bibinfo
  {year} {2015})\BibitemShut {NoStop}%
\end{thebibliography}
%

\end{document}